# Outflows from Massive YSOs as Seen with the Infrared Array Camera

Howard A. Smith, J. L. Hora, M. Marengo
*Harvard-Smithsonian Center for Astrophysics, 60 Garden Street, Cambridge, MA 02138*
hsmith@cfa.harvard.edu
and
Judith L. Pipher
*Department of Physics and Astronomy, University of Rochester, Rochester, NY 14627*

## ABSTRACT

The bipolar outflow from the massive star forming cluster in DR21 is one of the most powerful known, and in IRAC images the outflow stands out by virtue of its brightness at 4.5 μm (Band 2). Indeed, IRAC images of many galactic and extragalactic star formation regions feature prominent Band 2 morphologies. We have analyzed archival ISOSWS spectra of the DR21 outflow, and compare them to updated $H_2$ shocked and UV-excitation models. We find that $H_2$ line emission contributes about 50% of the flux of the IRAC bands at 3.6 μm, 4.5 μm, and 5.8 μm, and is a significant contributor to the 8.0 μm band as well, and confirm that the outflow contains multiple excitation mechanisms. Other potentially strong features, in particular Brα and CO emission, have been suggested as contributing to IRAC fluxes in outflows, but they are weak or absent in DR21; surprisingly, there also is no evidence for strong PAH emission. The results imply that IRAC images can be a powerful detector of, and diagnostic for, outflows caused by massive star formation activity in our galaxy, and in other galaxies as well. They also suggest that IRAC color-color diagnostic diagrams may need to take into account the possible influence of these strong emission lines. IRAC images of the general ISM in the region, away from the outflow, are in approximate but not precise agreement with theoretical models.

*Subject headings*: ISM: individual (DR21) – ISM: jets and outflows – ISM: molecules – infrared: ISM – stars: formation

*Online material*: color figures

1. INTRODUCTION

DR21 contains one of the most massive star formation regions and molecular outflows in our galaxy. It is an excellent test bed for studying the conditions of massive star formation, and for comparing competing ideas about the formation and subsequent evolution of massive new stars and their environments, for example, the relative roles of cloud collisions and fragmentation in these processes. The extended DR21 nebula consists of a filamentary ridge of massive star-forming clusters that stretch from DR21(main) northward to W75N, and includes clusters around the bright maser source DR21(OH). Along with the rest of the Cyg X complex in which it is embedded, DR21 appears as part of a shell-like structure around the super-massive Cyg OB2 association. This association lies in the direction of a tangent to the Orion arm of the Galaxy, and as a result its distance is relatively difficult to estimate. Dame, using CO and HI maps of the galaxy, estimates the distance to Cyg OB2 as 1.7kpc (private communication; Butt et al. 2003). We adopt this value for DR21 even though the most commonly used distance in the past literature has been 3 kpc. We thus revise downward a number of earlier estimates of luminosity, etc., so as to scale accurately to this closer distance.

The giant molecular outflow in DR21 is one of the most powerful outflows in our galaxy. An estimated driving force (estimated from the $H_2$ lines) of ~5.5 $10^{30}$ dynes (Garden et al. 1986) exceeds the driving force of the Orion-Irc2 outflow by almost a factor of 6 (Cabrit & Bertout 1992). The outflow velocity of about 60 kms$^{-1}$ or more drives a mass estimated at >3000Mo. The energy of the flow is in excess of $2 \times 10^{48}$ ergs; the luminosity of the two-micron $H_2$ flow alone has been calculated to be 1800 Lo, and, as we show below, the 2 μm lines are not even the brightest ones. Cyganowski, *et al.* (2003) claim evidence for a second "highly collimated" flow perpendicular to the main one. In part because of the distance to DR21, and also because of the high extinctions there, our understanding of the complex has been somewhat limited. That DR21 has regions of very high extinction was shown by Chandler *et al.* (1993a) and Chandler *et al.* (1993b), who use the 1.3-mm dust and CS line observations, respectively, to conclude that clumps exist with column densities as high as $N(H+H_2) \sim 3 \times 10^{24}$ cm$^{-2}$ ; $C^{18}O$ observations (Wilson & Mauersberger, 1990) had provided estimate that were also substantial but about ten times less. Adjusting to the closer distance of 1.7kpc and converting to visual extinctions give an estimate in places of Av~1000. *Spitzer* IRS data on the embedded young protostar finds a very deep $CO_2$ absorption also consistent with this estimate (Smith *et al.* 2006, in prep.). The IRAS satellite detected a very bright point source near the apparent source of the massive outflow, but identifying the actual source of the flow, a massive YSO, has proved elusive. Previous studies of the outflow region identified a source "IRS1" as the driver for the outflow, but IRAC images find no point source coincident with the IRAS location. Instead, we find one overwhelmingly strong 8.0 μm (Band 4) point source located about 10" from the position of the K-band star first dubbed "IRS1"(Davis &Smith 1996). This IRAC source was undetected at K; it is a young ~O7 star with $L_{FIR} \sim 1.5 \times 10^5$ Lo, apparently still accreting from its envelope; a full description of the source and its environment is in Smith *et al.* (*2006 op cit*). It is this source which is thought to power the immense bipolar outflow from the cloud.

The outflow has been imaged and studied in $H_2$ by numerous authors (e.g., Garden *et al.* 1986; Garden *et al.* 1990; Davis & Smith 1996; and Fernandes *et al.* 1997). Fernandes *et al.* conclude that the 2μm line ratios are best fit by a PDR model with FUV field in the range of 2<g<3 and a preshock density of n = $3 \times 10^3$ cm$^{-3}$, and with a small shock contribution. Smith *et al.* (1998; hereafter "SED") used the Infrared Space Observatory's Short Wavelength Spectrometer (ISOSWS) to analyze five lines between 4.9 μm and 17 μm in one bright section of the outflow. They conclude there are both J and C-type shocks at work in the region, and distinguish this gas from the uv-excited gas at 2 μm modeled by Fernandes *et al.*

In this paper we present new IRAC images of the outflow region which dramatically reveal at 4.5μm the structures seen by the 2μm images with much higher spatial resolution. The original IRAC observations of this region (Marston *et al.* 2004) were partially saturated on the bright point sources, with some effect on the neighboring outflow region pixels; our new observations were taken in HDR and subarray mode. We also present and analyze archival ISOSWS spectra of the DR21 outflow, material not published in the original SED paper; we report the fluxes on an additional four H$_2$ lines as well as the on the atomic hydrogen Brα recombination line, and limits to other line fluxes and features falling over nearly all of the wavelength coverage of the IRAC instrument. The results show that the *Spitzer* IRAC instrument is a superb probe of shocked structures because of the strength of the H$_2$ lines and the placement of the IRAC bandpasses.

## 2. OBSERVATIONS AND ANALYSIS

### 2.1 *IRAC Observations of the Outflow*

The original DR21 IRAC observations were taken as an Early Release Observation (ERO) and published in Marston *et al.* (2004). We repeated those observations in HDR and subarray modes in regions around the bright point source and the outflow that had suffered from saturation effects. (Please see the *Spitzer* Observer's Manual 6.0 for details of these observing modes, which read out the data in much shorter time intervals.) A full description of the point source and its nature will be presented elsewhere (Smith *et al.* 2006). The data analyzed here were processed in pipeline S11.0.2. Figure 1 shows the IRAC 3.6 µm / 4.5 µm /8.0 µm (Band 1,2,4) color composite of the western lobe of the outflow region.

Davis & Smith (1996) obtained high resolution, high SNR images of the $H_2$ 1-0 S(1) line emission from DR21. It was this same Davis and Smith paper which most convincingly identified a bright 2 µm star as the source of the outflow. As we show in more detail (Smith *op cit*) this is an erroneous identification, and the actual driving star is located about 15arsec away; it is not apparent in any K-band images, but dominates the entire region at 8 µm. Davis and Smith used a narrow-band (Δλ = 0.02 µm) filter centered at 2.12 µm to obtain their $H_2$ image, and also a wide-band K filter, and subtracted the scaled K-band image from the narrow-band image to produce the line map. Their spatial resolution was 0.63 arcsec/pixel. In Figure 2 we superimpose contours of their 1-0 S(1) image on our (lower resolution) IRAC 4.5 µm (Band 2) image. There is a nearly exact correspondence between the $H_2$ peaks and filaments and the 4.5 µm structures. This close correspondence across the ~1 parsec length of the flow suggests that the conclusions we draw from an analysis of the smaller ISOSWS field can be safely generalized to the entire shocked outflow. Superimposed on the image is the placement of the ISOSWS beam for the primary spectral observation, TDT 04402144 - it is this region of the shock which we analyze in detail below.

## 2.2 *Infrared Space Observatory's Short Wavelength Spectrometer Observations*

ISOSWS obtained seven sets of observations of the DR21 "West Lobe" in addition to another fifteen observations of varying quality centered elsewhere around the region. Wright *et al.* (1997) published an initial SWS analysis of the outflows in a short abstract, and concluded a combination of multiple shock components and some uv excitation were probably at work. SED analyzed and published the results from one of the outflow observations, TDT 34700904, in their meticulous analysis of the $H_2$ emission from DR21 West. That SWS observation, which was not a full wavelength scan but rather a set of individual line scans, covered and obtained data on five of the $H_2$ lines: the 0-0 S(1) at 17.03 μm, and four that fall in the IRAC coverage: the 0-0S(5) and 0-0S(7) lines, and the 1-1 S(7) and 1-1 S(9) lines. We reanalyzed that dataset, and are in agreement with all of their line fluxes except for the flux in the 0-0S(7) line, for which we measure a value 22% larger, more than the uncertainly in the data; we use our value below, and note that the difference may be due to our use of a more recent pipeline processing and calibration. Based on the observed line strengths, and combined with 2 μm observations of $H_2$, Smith, Eisloffel, and Davis concluded that the excitation in the DR21 outflow could not be produced by a simple C- or J-shock, and a range of shock strengths was needed. This conclusion was in contrast to the results of Fernandes, Brand, and Burton (1997), who relied on ground-based observations (only) to argue that the $H_2$ lines in the DR21 outflow were best fit by a PDR model with FUV field in the range of $2<g<3$ and pre-shocked density $n>3\times10^3\,cm^{-3}$. Smith & Rosen (2004) extended the Fernandes *et al.* analysis to simulate the *expected* images from IRAC observations under a wide range of shocked conditions.

We examined all seven sets of ISOSWS observations of the DR21 West lobe outflow in an effort to try to understand the IRAC images more precisely. None of the SWS scans provided full, contiguous coverage. TDT 04402042 emphasized wavelengths over 10 μm, TDT 04402144 (Figure 3 shows this spectral scan in the 7 – 9.5 μm interval) was contiguous from about 6.5 to 9.5 μm, and TDT 19301741 consisted of a series of some twenty-one short line scans between 2.2 μm and 36 μm. All these TDTs were taken at different epochs and included slightly different portions of the brightest part of the outflow, but all were centered roughly on the same part of the outflow, differing in location only by about 10" ( the latter TDT also used a larger beamsize, 20"x33" ). The new observation we use most, TDT 04402144, was centered 9.4 arcsec to the NE of the SED slit, and was rotated clockwise by 56 degrees. None of these differences have any substantial effect on our rather general conclusions. The new TDTs we analyze here provide high SNR fluxes on four *additional* $H_2$ lines in the IRAC wavelength coverage beyond the five analyzed in the SED study: the 0-0 S(9) and 0-0 S(11) lines, the 1-1 S(5) line, and the 1-0 O(5) line. In addition, SWS detected the weak hydrogen recombination line Brα line at 4.052 μm, and set useful limits on some others.

### 2.3  *The Strong $H_2$ Features*

The SWS spectral scan of the bright outflow over the 7 - 8.8 μm interval is presented in Figure 3. The $H_2$ 0-0 S(4) and S(5) lines clearly dominate this interval. For all of the line we calculate their corresponding fluxes in IRAC by converting each of the observed line strengths into a flux density in the appropriate IRAC band using the IRAC instrumental response procedures as described in the IRAC Data Handbook (Section 5.2). The Handbook Version 2.0 contains some significant errors in the Photometry and Calibration section (Chapter 5). In particular, the expression for converting flux into flux density (page 41) is missing a wavelength ratio and should read: $F = F_{\nu o \text{ (quot)}} \times \Delta \nu \times (\lambda/\lambda_o)/R_1$, and the Handbook Table 5.2 lists inaccurate values for the bandwidth $\Delta \nu$ to be used in this expression; they are about 30-40% too high in the Version 2.0 table. We have used the corrected values (B. Reach, private communication) as will appear in the new version of the Handbook (Version 3.0). The flux density in the $H_2$ 0-0 S(4) line, converted to what it contributes to the IRAC 8.0 μm band, is 38.0mJy. The 0-0 S(5) line at 6.907μm contributes another 66.0mJy, making a total line flux density in the 8.0 μm band of 104 " 10 mJy. All of the observed $H_2$ lines detected by ISOSWS observations are listed in Table 1.

The results confirm that in the outflow region of DR21 the $H_2$ lines contribute significantly to the IRAC band fluxes. We find that in the 4.5 μm, 5.8 μm, and 8.0 μm bands the ISOSWS *observed* line strengths alone account for between 13% - 30% of the total measured IRAC fluxes in the outflow region studied by SWS (as indicated on Figure 2); this value drops to about 6% for the 3.6 μm band. The slight uncertainties in the respective ISO measurements, for example due to the slight rotations in the SWS beam between the different TDT observations of different lines, do not appreciably alter any of our conclusions. Figure 4 plots the measured IRAC flux densities summed over the region equivalent to the ISOSWS beam (filled diamonds), along with the contributions from the observed ISOSWS $H_2$ lines in those IRAC bands (filled diamonds).

## 2.4 *The Faint Continuum and PAH Features*

Particularly worthy of notice in the mid-IR spectrum shown in Figure 3 (upper) is the extremely faint continuum on which the $H_2$ lines sit, only about 0.93 " 0.06 Jy. The complete absence of any characteristic sign of PAH emission features in the spectrum is striking. For comparison, Figure 3 (lower) shows the spectrum of the bright region in DR21 around the driving source about 40arsec away. Here the strong PAH features completely dominate the spectrum, and the $H_2$ 0-0S(4) line, although present, is a peripheral feature. We were able to investigate the strength of the 3.4 μm PAH feature in the outflow as well, since two of the seven sets of SWS scans of the outflow did have marginally useable subscans over the this feature. In one scan (TDT 04402144) we were just able to detected the presence of this feature at 3σ, at a level of 1.2 +- 0.4 $x10^{-19}$ watts/$cm^2$; it was undetectable in the others due to high noise. Figure 3 (lower) of the main DR21 cluster also shows clearly the emission lines from [ArII] 6.985 μm, [ArIII] 8.991 μm, and the HeII 12-10 line at 7.46 μm, that are excited by the embedded O star (Smith *et al., op cit*).

The absence of PAH emission in the outflow was a surprise. It is well known that PAH features are weak or absent in the immediate vicinity of active galactic nuclei (e.g., Risaliti, 2004; Lutz, 2000), in some compact HII regions (Roelfsema *et al.* 1996), and in some planetary nebulae (Bernard-Salas &Tielens, 2005). The reasons for these differences in PAH emission are not well understood. Strong ultraviolet radiation (energies over about 50eV) will destroy PAHs (e.g., Risaliti, op cit), for example in the region of AGN; dehydrogenated PAH species have weaker C-H bands (the 3.4 μm and 11.2 μm in particular; Pauzat et al. 1997; Verstrasete et al 2001); and large PAHs ($N_{carbon} > 100$) have dominant 6-16 μm features while small ones dominate at 3.3 μm (Schutte et al., 1993). The ionization state is also a key parameter for PAHs. Huggins & Allamandola (2004) provide extensive laboratory data that indicate the 5.8 μm and 8.0 μm band PAH features are comparatively very weak in neutral species, but are strong in ionized PAHs. The example here of DR21 West shows that PAH emission can change dramatically over only a few tenths of a parsec, as the environment changes from the neighborhood of a hot young massive star to the outflow of a rapidly moving shock. While it is clear that several types of processes are able to destroy or modify the PAH molecules in a way consistent with our observations, we cannot tell from our data alone which of these potential scenarios is the correct explanation. We will return to this general problem in the context of the general ISM in Hora *et al.* (2006). For completeness, we note that the zodiacal light contribution to the four bands at the epochs of observation is small; it is estimated by *Spitzer* as being 0.2mJy, 1.0mJy, 9.3mJy, and 26.9mJy respectively.

## 2.5  The Unseen Features: Br α and CO

IRAC images of star formation regions and other nebulosity-rich objects often reveal strong 4.5μm band emission, which is easily spotted as bright green nebulosity in 3-color IRAC images in which the 4.5 μm band is coded (typically) as green. Churchwell *et al.* (2004), in their analysis of reflection nebula RCW 49, see just such bright 4.5 μm band emission. They conclude that the atomic hydrogen Brα line at 4.052 μm provides about 20% of the nonstellar flux in the 4.5 μm band , dominating the extended emission in that source. In the case of DR21, however, we measure the Brα line with SWS to be 1.6 " 0.4 x$10^{-16}$ Watts/m$^2$, making it among the weakest of the emission lines in the outflow within the 4.5 µm band window, and 10 times weaker than the H$_2$ 0-0 S(9) line. No other hydrogen recombination lines are seen in the SWS spectra of the outflow lobe. We conclude that "green nebulosity," insofar as it resembles shocked outflows like DR21, signifies the presence of H$_2$ , not Brα.

Geballe & Garden (1990) used UKIRT to obtain spectra of the Orion shocked outflow at 4.74µm, the wavelength of the CO 1-0 P(8) line. They find it has a strength about 1/10 of the strength of the H$_2$ 0-0 S(9) line there, and conclude, in agreement with some earlier observers, that shocked CO is an important contributor to the emission in the Orion outflow. Van Dishoeck *et al.* (1998) publish an ISOSWS scan of the Orion IRc2 region, which includes the region studied by Geballe & Garden; they report "possibly" seeing CO emission in the 4.4 - 4.8 µm region. In the ISOSWS scans of the DR21 West outflow, however, no CO lines are detected to a limit of about 3x$10^{-16}$ Watts/m$^2$ – less than about 16% of the nearby H$_2$ 0-0 S(9) line.

### 3. MODELS of the H$_2$ EMISSION

#### 3.1 *Shock and PDR models of the Outflow*

There are about 140 emission lines of H$_2$ in the IRAC bands whose strengths, depending on the model details, are within 1% of the strength of the strongest line in each excitation case. Michael Kaufman and Mark Wolfire have provided us with their as yet unpublished results of updated PDR and shock models (Kaufman *et al.* 2005; see also Kaufman & Neufeld, 1996, and Wolfire *et al.* 1990). Figure 5 plots the brightest $H_2$ lines in the IRAC bands for the case of a preshocked density of $1 \times 10^4$ cm$^{-3}$ moving with one of three different velocities, 20 kms$^{-1}$ (asterisks), 30 kms$^{-1}$ (triangles), and 40 kms$^{-1}$ (diamonds). As might be expected, the strongest velocities produce by far the strongest line emission. It is also notable that the strongest velocities have the *least* flux variation in line fluxes across the IRAC bands, whereas the weaker velocities produce emission lines whose contributions between the IRAC bands vary much more substantively.

Fernandes *et al.* (1997), in their earlier analyses of ground-based observations of the DR21 outflow, concluded that the DR21 $H_2$ line ratios were best fit by PDR models because of the strength of the lines originating from the upper state v=2, 3, and 4 levels; these are preferentially populated via non-thermal mechanisms, although there they also found a low-excitation shocked component. Figure 6 plots the brightest 90% of the bright $H_2$ lines predicted by Wolfire and Kaufman in the IRAC bands for five different cases of density n and ultraviolet flux g (where g is the log of the flux in units of the local interstellar FUV radiation, taken as $2 \times 10^{11}$ photons/sec-m$^2$). The strongest emission obviously comes from the most dense, highest uv regions, with the 8.0 μm band encompassing the brightest lines and the 3.6 μm band the faintest ones. However, it is noteworthy that there are many more detectable lines in the shorter bands, whose cumulative contributions can be substantial.

The four new $H_2$ lines in DR21 we present here, the 0-0 S(9), 0-0 S(11), 1-1 S(5), and 1-0 O(5) lines, and the limits we can set to some lines *not* seen, allow us to make additional statements about the excitation mechanisms. Fernandes *et al.* concluded from their data that the most likely dominant excitation in DR21 is from a FUV field with $2 \# g \# 3$ and a preshock density of $n \$ 3 \times 10^3$ cm$^{-3}$. The PDR model most analogous to the model parameters we have (n=3, g=3) predicts in that case a strength for the 4-3 O(3) line at 3.3765 µm (one of the strongest discriminating H$_2$ lines) of $1.0 \times 10^{-18}$ Watts/m$^2$. We set a limit from our analysis of the ISOSWS spectrum that is about 50 times fainter, thereby ruling out this case. Higher densities or uv flux values will reduce the predicted line flux ratio, and in the case of n=5 and g=4 the 4-3 O(3) line is about 150 times weaker, consistent with the new observational limit. However, in this case the longer wavelength lines, including the v=0 pure rotational series, become very bright. We conclude that the most likely average PDR situation is more consistent with an n=4 g=4, and we take this as our baseline UV case for modeling. The new lines also add some additional consistency to the shock models. SED concluded a mixture of C and J shocks were present. For our purposes, a shock with density of $10^4$ cm$^{-3}$ and a velocity of 30 kms$^{-1}$ gives reasonable fits. In the more rapid shocks the relative contribution of H$_2$ lines to the 4.5 µm band lines is increasingly dominant, in part because this band (and the 3.6 µm band) include a considerably larger number of weaker lines which are excited more effectively only at higher velocities, as was seen in the UV-excited scenario. But the higher velocity shocks are also intrinsically brighter.

The contribution of the *observed* H$_2$ lines to the IRAC band flux densities are listed in Table 1. We wanted to estimate the total contribution of the H$_2$ emission to the IRAC bands, because the ISOSWS scans were relatively sparse, in Bands 1 and 2 especially. Our IRAC image (Figures 1, as well as the 2 µm image (Figure 2), show that the shocked H$_2$ emission in the outflow encompassed by the beam of ISOSWS includes many small knots. The calculated H$_2$ line brightness tables by Kaufman must therefore be corrected by appropriate, and somewhat uncertain, dilution factors when comparing absolute model flux predictions to the SWS observations presented in the tables. To make our estimate, we therefore normalized a shock model with density of $10^4$ cm$^{-3}$ and a velocity of 30 kms$^{-1}$ to the strong, observed 0-0S(4) line, and we scaled a n=4 g=4 UV-excitation component to the strong, observed 1-1 S(7) line – a line which is very weak in all shocked models. We then calculated the contribution of all of the $H_2$ lines to the IRAC bands from this combined model. The results are plotted in Figure 4 as crosses. As expected, the modeled lines add considerable flux density, especially in the 3.6 μm band. When taking into account the $H_2$ lines that were not observed directly by ISOSWS, the $H_2$ lines contribute about 50% of the total flux in IRAC 3.6 μm, 4.5 μm, and 5.8 μm bands; in the 8.0 μm band the lines contribute about 20% of the flux.

Some of the individual, observed line fluxes do not match the intensities predicted by this simple, two component, normalized model, especially in the short wavelength band. Adding other C-shock components from higher density regions could provide a solution, since at higher densities (as noted earlier) it is precisely the 3.6 μm, 4.5 μm fluxes which are most increased. SED concluded that it was not possible to use a single shock model to explain the ISOSWS and ground-based lines they measured. Instead, they propose that a combination of C-type bow shocks with "wide flanks," together with localized extinction, can provide a consistent picture of the region. Alternatively, some component of low density PDR emission might accomplish the same thing, but would require a larger filling factor and a more careful analysis to explain the absence of the 4-3 O(3) line. Our own data do not allow us to say much more the details of modeling, but neither will adjustments alter our main conclusion: that the $H_2$ lines play a major role in determining the IRAC fluxes and consequently the colors of the IRAC images.

### 3.2 *The Character of the ISM in DR21 as Determined by IRAC*

Up to now we have analyzed the outflow in the region of the observed ISOSWS beam treated as an averaged whole. There is every reason to suspect, however, that additional details of the shock's structure can be revealed in the relative IRAC colors of the various knots and filaments. Figure 7 shows an IRAC color-color plot with all of the pixels in the giant outflow (green points) and in the surrounding, unshocked nebulosity (blue crosses), given in magnitudes in order to be most useful for comparison with point-source colors. The uncertainties in the colors of the fainter, diffuse nebulosity are larger than those in the bright knots, but do not affect our conclusions.

The colors of the emission of the outflow itself form a narrow locus of points stretching from a [3.6 - 4.5] color of 1.5mag in the top left of the color-color plot, down to a [5.8 – 8.0] color of 1.8 in the lower center. This locus is roughly the same as that produced by shocked $H_2$ lines alone, with preshock density of $1 \times 10^4$ cm$^{-3}$, when the range of velocities varies from 20 - 40 kms$^{-1}$; the top-most left portion corresponds to those lines originating from the fastest material. When the outflow is subdivided in smaller regions and their colors are separately considered, it is apparent that this range of colors is roughly the same in each zone. It is already clear from the work of SED that multiple shock components are present; the variable colors across the flow simply confirm that complex and changing conditions are present.

There is one region in the outflow that is different from the others: the faintest part of the outflow – viz., the region at the western tip – is lacking in the reddest [3.6 - 4.5] gas, and instead has gas whose color lies only between about 0.6 and 1.2 mag. We do not have spectra of the western tip, but we note that decreasing the preshock velocity by about 10 kms$^{-1}$ will decrease the cumulative line flux ratio in these bands in the most probable, low density case, by a factor of ~5. It does so largely by reducing the 4.5 μm band line emission. Although the line contributions to this band and the 3.6 μm band are only about 50% of the total flux in the bands, the changes will bring the color closer to that observed. Increasing the densities by a factor of ten can accomplish about the same thing in the higher velocity cases.

The background nebulosity, away from the massive strong outflow, is also of considerable interest, and we have also examined its IRAC colors. Figure 7 plots the colors of the non-outflow nebulosity. The IRAC images away from the outflow are obviously less "green" in appearance when weighed with the customary 3-band combinations, meaning less intense H$_2$ emission in the 4.6 μm band. Since the total flux densities are also much smaller away from the flow, we have averaged these nebular regions over much larger areas than were used in the outflow itself. All of the non-outflow, nebulous regions are characterized by very modest reddening in the 3.6 μm and 4.5 μm bands -- typically a color less than [3.6 - 4.5] < 0.4mag; this is in contrast to values of 1.0 or greater in the outflow region. Red [3.6 - 4.5] colors are as strong an indicator of shock activity as is strong 4.5 μm emission. In contract, the [5.8 – 8.0] colors of the diffuse nebulosity are virtually indistinguishable from the average colors in the outflow itself. Within the diffuse material there is a slight tendency for faint material to be even redder than brighter nebulosity, perhaps suggesting that uv, which excites the PAH emission there, plays a more dominant role in these regions. Li and Draine (2001) provide detailed quantitative models for the ISM in which they include PAH, silicate grain, and carbonaceous grain emission. Their Table 5 provides estimated, normalized flux densities for a range of seven values of the UV radiation field strength, $\chi_{MMP}$, between 0.3 and $10^4$. The IRAC colors of all these seven scenarios are about the same: $[3.6 - 4.5] \approx .35$; $[5.8 – 8.0] \approx 2.1$. The points are considerably bluer in $[3.6 - 4.5]$, by a factor of at least 60%, than all but the bluest of nebular regions in our image; the band $[5.8 – 8.0]$ color is about 0.25mag redder than we observe. A more complete discussion of the IRAC properties of the ISM, including a more detailed analysis of the ISM in the DR21 region, will be presented in Hora *et al.* (2006, in prep.).

## 4. CONCLUSIONS

*Spitzer* IRAC color images of star formation regions, typically coded with the 4.5 μm band as green, often show dramatic swaths of green nebulosity. In the case of the massive outflow in DR21, that "green monster" is seen largely thanks to the strong lines of shocked or UV-pumped $H_2$. Because the 4.5 μm band lacks strong PAH features such as can dominate the 5.8 μm and 8.0 μm bands, when composite IRAC images are scaled down (and corrected for bright point sources) so that the bright PAH regions do not overwhelm the image, the outflows show up as green nebulosity.

The ISOSWS spectra of the DR21 western outflow lobe have allowed us to determine not only that $H_2$ is the overwhelming contributor to the outflow emission lines, they also enable us (in agreement with earlier analyses) to conclude that some combination of modest velocity C-shocks in a medium whose preshock density is between $10^3$ and $10^4$ cm$^{-3}$, coupled with some UV-excitation in a region of similar density and a UV field characterized by g=4, are the mechanisms responsible. We have not only found that the $H_2$ emission in the outflow is strong – we have found that the continuum emission is very weak. The implications, to be discussed further in the larger context of the DR21 star formation complex and sources (Smith *et al.* 2006), is that neither dust nor hot PAH molecules are abundant in this outflow. The ISOSWS spectra of the outflow contain no strong PAH features, no CO bandhead emission, and only very weak Brα emission.

IRAC color-color diagrams of the emission from regions across the outflow confirm the conclusions from the individual band flux analyses. In particular, the short wavelength colors in the flow are consistently redder than are the colors of the non-outflow, non-shocked ISM. Combined with shock models, the colors indicate that the shock at the western tip is likely to involve slower, denser material. The colors of the background nebulosity, away from the massive strong outflow, are characterized by very modest reddening between the 3.6 μm and 4.5 μm bands. They are mild disagreement with theoretical predictions of for ISM colors in both [3.6 - 4.5] and [5.8 – 8.0] colors.

IRAC images have made it relatively easy for scientists to spot small regions of the 4.5 μm band activity in otherwise large and complex fields. While the *Spitzer* IRS spectrometer cannot see the $H_2$ lines below 5 μm, its sensitive performance ought to enable a self-consistent picture to emerge from the other, longer wavelength $H_2$ lines that do fall in its windows. Both the strength of the 4.5 μm band, and the [3.6 -4.5] color, are useful diagnostics of $H_2$ emission. The overall results indicate that IRAC images can be a powerful indicator of outflows caused by star formation activity. The results also suggest that IRAC color-color diagnostic diagrams for the nebulosity in star formation regions may sometimes need to take into account the possible influence of these strong $H_2$ emission lines, especially in cases when a large, extended emission is seen around an optically thick point source. It remains to be determined the extent to which the ~140 $H_2$ emission lines across the IRAC bands contribute to the IRAC images of other star formation regions (or for that matter of other bright nebulae, like planetaries, objects which are already known to be strong $H_2$ emitters).

We gratefully thank Mark Wolfire and Michael Kaufman for providing us with their latest calculations of the $H_2$ line strengths. We also thank Chris Davis for providing unpublished data on his 2 μm $H_2$ images. HAS acknowledges partial support from NASA Grant NAG5-10654. This work is based in part on observations made with the *Spitzer* Space Telescope, which is operated by the Jet Propulsion Laboratory, California Institute of Technology under NASA contract 1407. Support for the IRAC instrument was provided by NASA through Contract Number 960541 issued by JPL. This work is also based in part on observations made with ISO, an ESA project with the participation of NASA and ISAS.

Table 1

Fluxes of the ISOSWS -observed H₂ Lines

| Line ID | wavelength | Observed flux | flux contribution to IRAC bands |
|---|---|---|---|
|  | μm | (x$10^{-15}$ W/m^2) | (mJy) |
| 0-0S(4) | 8.0250 | 4.80 " .16 | 38.0 |
| 0-0S(5) | 6.9091 | 11.5 " 1.6 | 66.0 |
| 1-1S(7) | 5.8111 | 0.13 " .02 | 1.1 |
| 0-0S(7) | 5.5115 | 6.50 " .19 | 45.5 |
| 1-1S(9) | 4.9533 | 0.15 " .02 | 0.91 |
| 0-0S(9) | 4.6947 | 1.81 " .08 | 13.0 |
| 0-0S(11) | 4.181 | 1.18 " .2 | 7.2 |
| 1-0 O(5) | 3.235 | 0.54 " .04 | 2.6 |

Table 2

IRAC fluxes

| | | IRAC fluxes[1] | Observed $H_2$ lines[2] | Modeled $H_2$ lines[3] |
|---|---|---|---|---|
| IRAC band | wavelength | (mJy) | (mJy) | (mJy) |
| Band 1 | 3.6 μm | 46 | 2.6 | 25 |
| Band 2 | 4.5 μm | 71 | 21.1 | 34 |
| Band 3 | 5.8 μm | 320 | 46.6 | 123 |
| Band 4 | 8.0 μm | 760 | 104.0 | 151 |

(1) IRAC fluxes in a beam corresponding to the ISOSWS beam (14"x20")

(2) $H_2$ lines observed by ISO SWS, and converted to flux densities in the IRAC beam as described in the text.

(3) The sum of all of the $H_2$ lines in the IRAC bands from two models, one shock and one uv-excited, as described in the text.

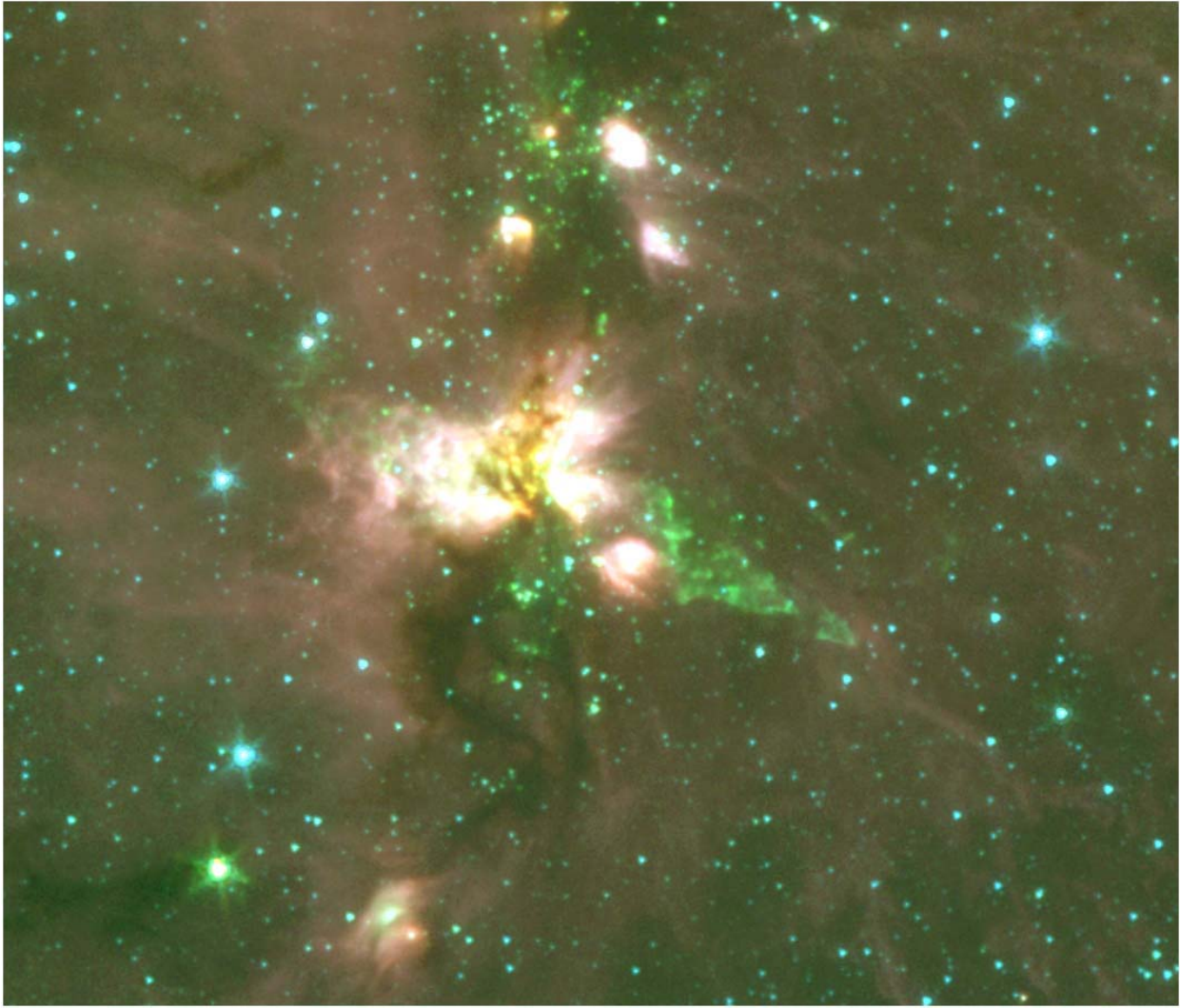

Figure 1: The IRAC Band 1 (3.6 μm - blue), Band 2(4.5 μm - green), Band 4( 8.0 μm - red) composite image of DR21-main.

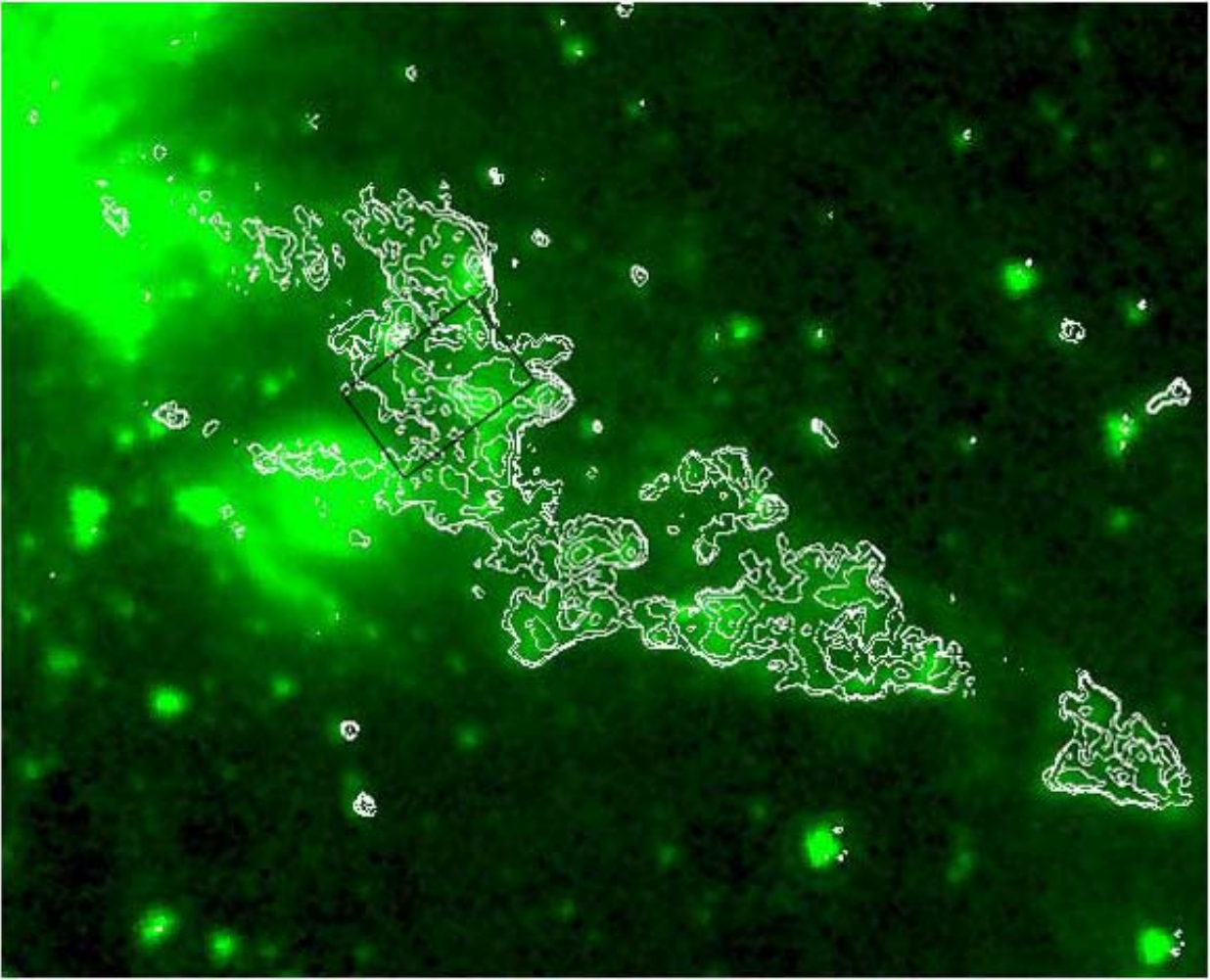

Figure 2: The IRAC 4.5 μm band image, with the K-band H$_2$ 1-2 S(1) line emission map overlaid in contours (from Chris Davis, private comm.) There is a very close correspondence between the H$_2$ structures and the 4.5 μm band knots. The ISOSWS 14"x20" field of view is outlined in black.

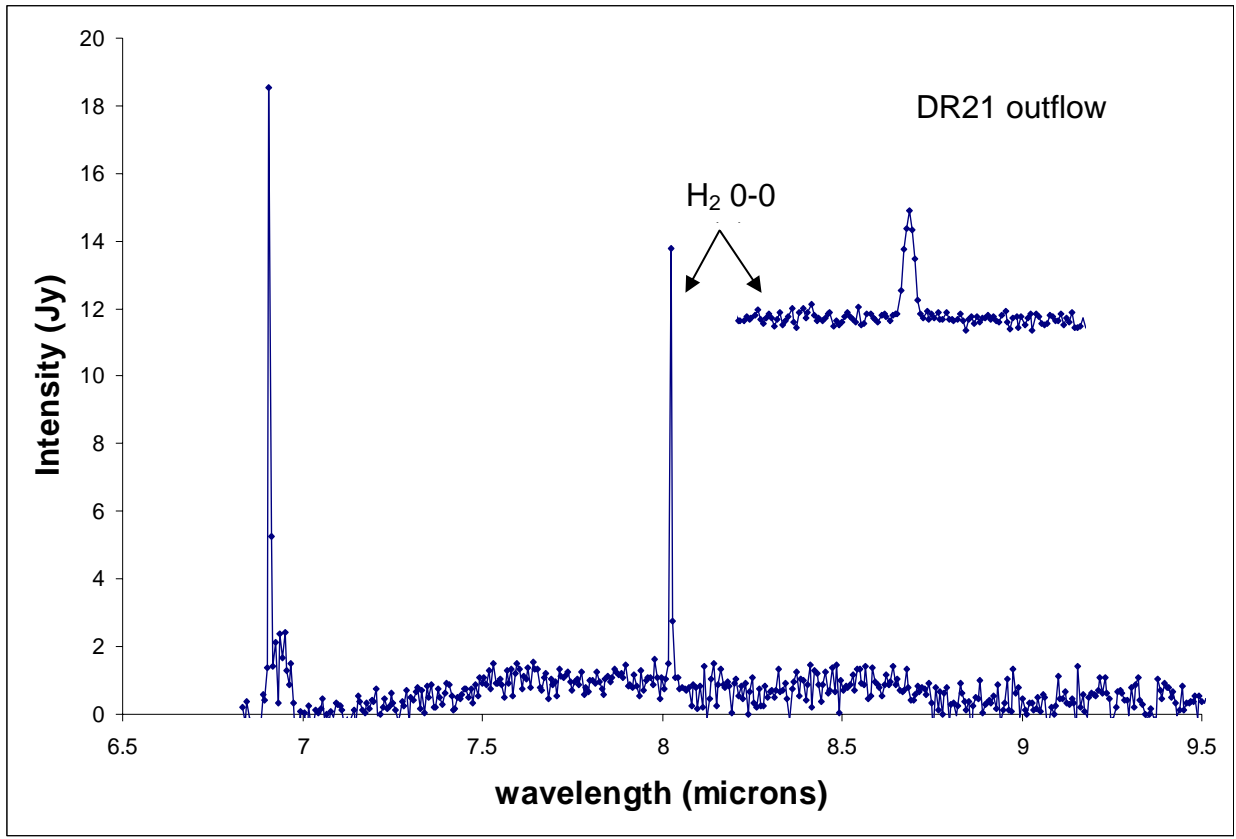

Figure 3 (upper): A portion of the ISOSWS scan of the outflow. It shows the 0-0 S(4) emission line at 8.025 μm; the inset shows a zoom on this line. Note the very low continuum emission in the outflow.

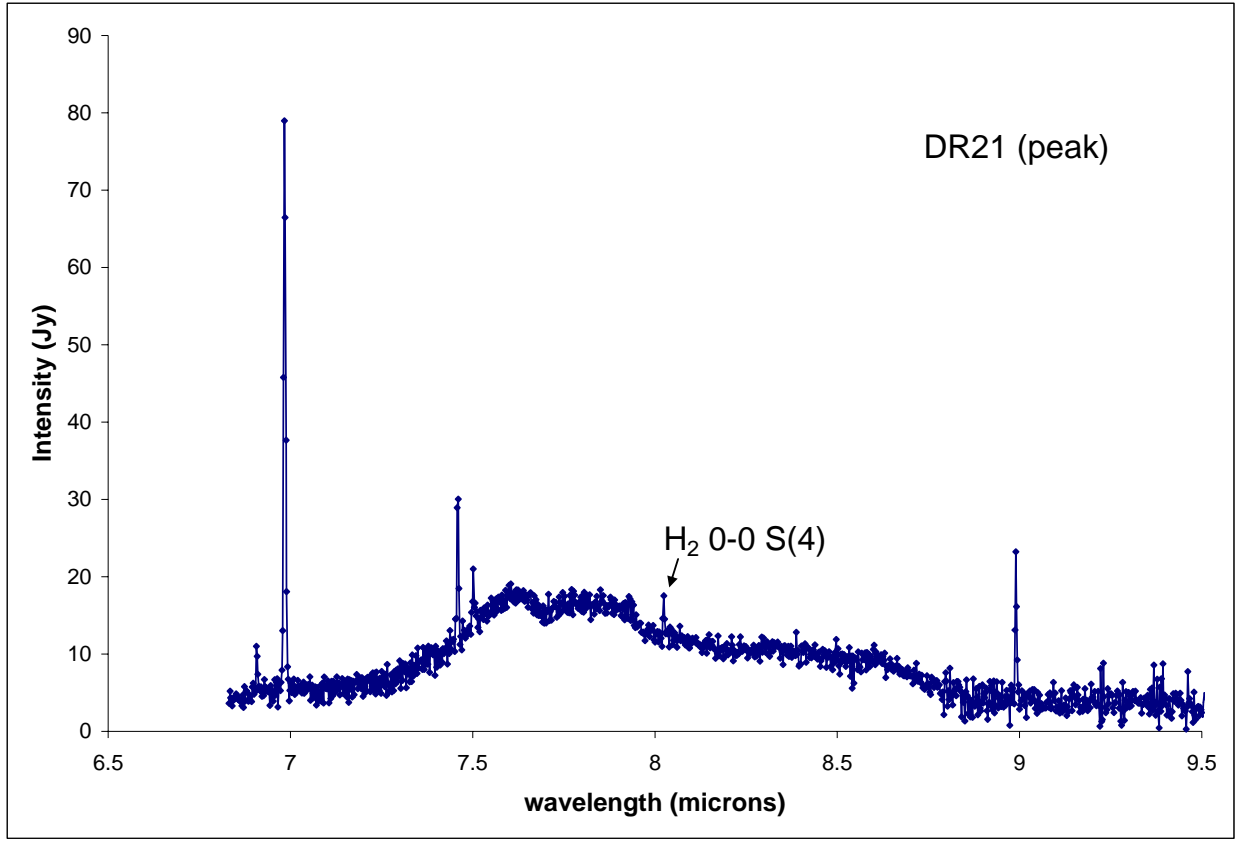

Figure 3 (lower): In dramatic contract to (3 upper), the ISOSWS spectrum of DR21 at the *peak* of the dust cloud shows strong continuum and PAH features, as well as lines of $H_2$, and atomic lines. The 0-0 S(4) line is visible, but is not at all prominent.

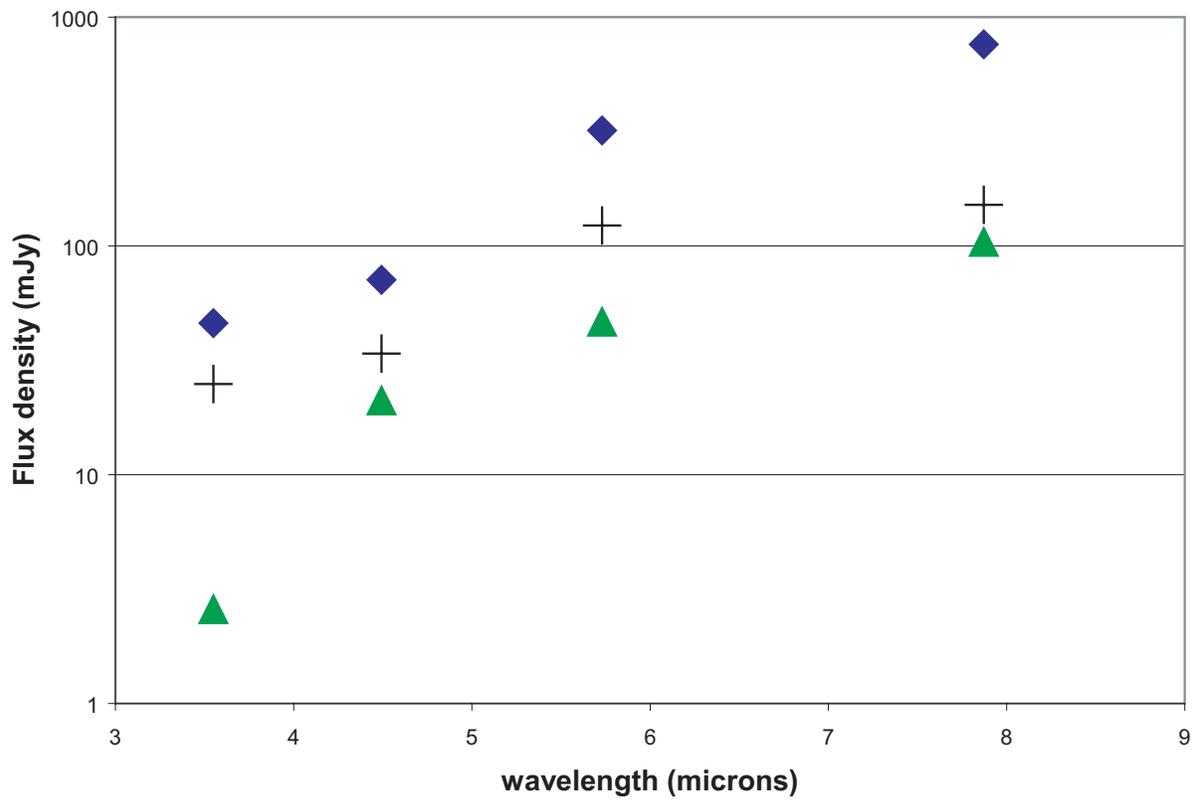

Figure 4: Comparison of the measured and modeled flux densities. The observed IRAC flux densities in the DR21 outflow, measured in the same beam as the ISO SWS spectral observations, are plotted as diamonds; the observed SWS line fluxes per band, converted to IRAC flux densities, are plotted as triangles. The pluses show the fit of a combined shock plus UV excitation model; each of these was normalized to one strong, observed line as described in the text; the residual flux is from the continuum. The $3\sigma$ statistical uncertainties for all of the observations are approximately the same sizes as the data markers.

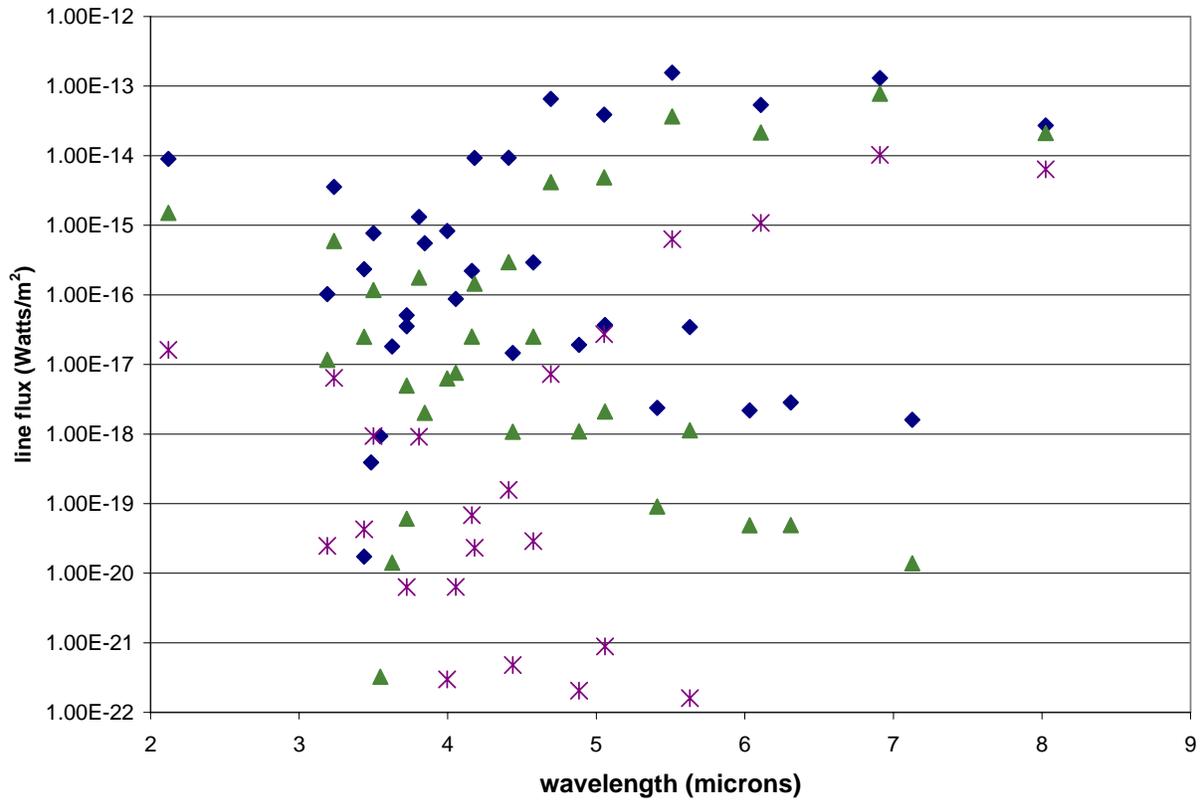

Figure 5: For the case of a preshocked density of $10^4$ cm$^{-3}$, a plot of the brightest shocked lines in the IRAC bandpasses versus wavelength for shock velocities of 40 kms$^{-1}$ (diamonds), 30 kms$^{-1}$ (triangles), and 20 kms$^{-1}$ (asterisks); Kaufmann (private communication)..

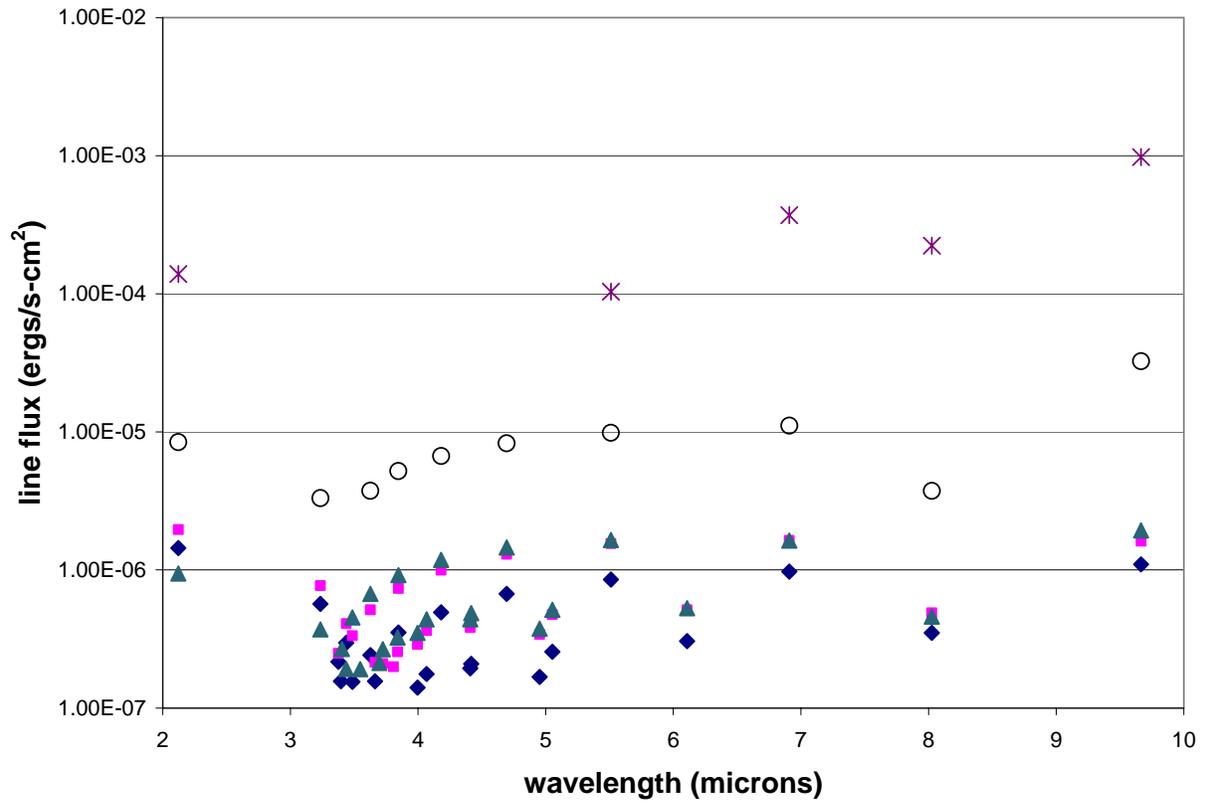

Figure 6: The brightest 90% of the H$_2$ lines in the PDR excitation models (Wolfire, private comm.) The five models shown are n=5, g=4 (asterisks), n=4, g=4 (open circles), n=3, g=5 (triangles), n=3, g=3 (squares), and n-3, g=2 (diamonds).

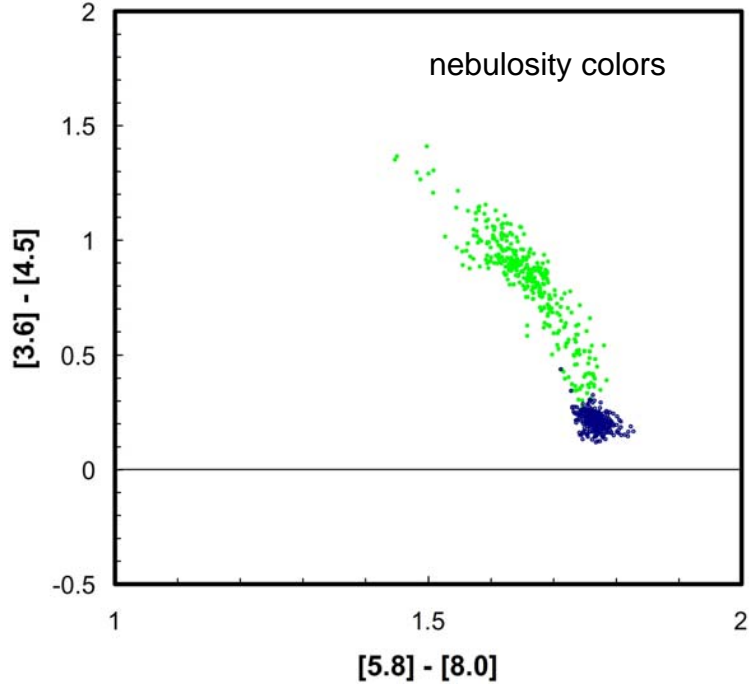

Figure 7: IRAC colors of the shocked outflow nebulosity (green/grey points) and surrounding nebulosity (blue/black circles). The points correspond to 2x2 averaged IRAC pixels across the outflow. Analysis by subregion finds that all parts of the flow except one have roughly the same range of pixel colors; the western-most tip is lacking the red [3.6]-[4.5] points with values above 0.6. The blue crosses correspond to pixels away from the outflow, in the general DR21 nebulosity. The expected IRAC colors of diffuse nebulosity from the Li and Draine (2001) ISM models fall in the plot around the value (2.1, -0.35), appreciably removed from the observed locus in DR21.